\documentstyle[aps,eqsecnum,epsf]{revtex}

\newcommand{\be}{\begin{equation}}
\newcommand{\ee}{\end{equation}}
\newcommand{\bea}{\begin{eqnarray}}
\newcommand{\eea}{\end{eqnarray}}
\def\b2c{\begin{multicols}{2}}
\def\e2c{\end{multicols}}

\begin{document}
\draft
\title{Renormalizability and the scalar field}
\author{Ramchander R. Sastry}
\address{Center for Particle Physics,\\
 University of Texas at Austin, \\
   Austin, Texas 78712-1081.}
\date{\today}
\maketitle

\begin{abstract}
The infinite dimensional generalization of the quantum mechanics
of extended objects, namely, the quantum field theory of extended
objects is presented.  The paradigm example studied in this paper
is the Euclidean scalar field with a $\lambda{\phi^{6}}/{ 6!}$
interaction in four spacetime dimensions.  The theory is found to
be finite when the virtual particle intermediate states are
characterized by fuzzy particles instead of ordinary pointlike
particles.  Causality, Lorentz invariance, and unitarity (verified
up to fourth order in the coupling constant) are preserved in the
theory.  In addition, the Kallen-Lehmann spectral representation
for the propagator is discussed.
\end{abstract}

\section{Introduction}
The necessity to incorporate the finite extent or fuzziness of a
quantum particle into our description of the world and the
resulting quantum mechanics have been presented by the
author\cite{sastry}.  In the extended object formulation we have
noncanonical commutation relations of the form
\be
\left[ X_{f_i},P_j\right] = ie^{-P^2/m^2}\delta_{ij}, \mbox{ and }
\left[X_{f_0},P_0\right] = -ie^{-P_0^2/m^2},
\ee
where $X_f$ is the noncommuting position operator and $P^2 =
P_iP^i$.  In terms of Euclidean momenta we can write the
commutation relations as
\be
\label{co-eqn}
\left[X_{f_{\mu}},P_{\nu}\right] = ie^{-P^2/m^2}g_{\mu\nu},
\ee
where $P_{\nu}$ is the Euclidean 4-momentum, and $g_{\mu\nu} =
(1,1,1,1)$ is the Euclidean metric and $P^2$ is a Euclidean
scalar.  The commutation relations in Eq.~(\ref{co-eqn}) have a
smooth limit with ordinary quantum mechanics as the Compton
wavelength ${1}/{m}$ of the particle vanishes.  Our goal is to
understand the infinite dimensional generalization of the quantum
mechanics based on these commutation relations, namely, the
quantum field theory of extended objects.  The characterization of
intermediate states by fuzzy particles obeying the commutation
relations in Eq.~(\ref{co-eqn}) becomes important in rendering
finite, hitherto nonrenormalizable theories because the
troublesome vertex which causes these divergences is effectively
smeared out.  In this paper we restrict ourselves to the simple
case of the Euclidean scalar field with a $\lambda{\phi^{6}}/{6!}$
interaction  which has been found to be nonrenormalizable when
approached via the quantum field theory of point particles.  We
choose to construct the field theory in Euclidean space for
convenience.  The generalization of the commutation relations in
Eq.~(\ref{co-eqn}) to continuous systems (fields) and the
subsequent development of the theory leads to a Euclidean momentum
space propagator which can render the interacting field theory
finite to all orders.  The theory is found to obey causality and
Lorentz invariance, and unitarity is demonstrated up to fourth
order in the coupling constant.  In addition, the Kallen-Lehmann
spectral representation for the propagator is presented.  We note
that Lorentz invariance of such quantities as propagation
amplitudes and scattering amplitudes is ensured when we employ a
Euclidean formulation by virtue of the variable transformation
$p_0 \rightarrow ip_0$. This cannot be viewed as a rotation of the
contour in the sense of the Wick rotation (since the contour does
not close), and as we shall demonstrate such a transformation
recovers the Minkowski space propagation amplitude.  Thus,
preservation of the four dimensional rotational symmetry of
Euclidean space will also preserve Lorentz invariance.

\section{The Scalar Field}
Super-renormalizable and renormalizable quantum field theories are
characterized by the dimensionality of their coupling constants
which have positive mass dimension or are dimensionless.  In the
first case, the perturbation expansion of the N-point functions or
scattering amplitudes have positive length dimension and the
degree of divergence of the graphs becomes smaller as more
vertices are included.  In renormalizable theories, the
perturbation expansions are dimensionless and only a finite number
of amplitudes superficially diverge; however, divergences occur at
all orders in perturbation theory and can be automatically removed
by a redefinition of a finite number of physical constants.
Nonrenormalizable theories are characterized by negative
dimensional coupling constants.  Consequently, the perturbation
expansions have dimension $\frac{1}{length}$ and as we include
more vertices the degree of divergence becomes larger.

A point particle renormalizable theory can be rendered finite by
implementing regularization procedures; for example, by
introducing a high momentum cut off $\Lambda$ in which case we
renounce all interest in regions less than $\Lambda^{-1}$ and
consequently, the point of interaction (the vertex) is fuzzed out.
However, such procedures fail to render nonrenormalizable theories
finite while preserving essential features such as gauge
invariance, Lorentz invariance, and unitarity.  Instead of
introducing such cut offs we can take advantage of the fuzziness
inherent in the particle to place an effective lower bound on the
short distances.  It is for this reason that we introduce fuzzy
particle intermediate states in a nonrenormalizable theory.

The Euclidean space Lagrangian for the scalar field with a
$\lambda \phi^6/6! $ interaction is written as
\be
{\cal L} = \frac{1}{2}(\partial_{\mu}\phi)^{2} +
\frac{1}{2}m^{2}\phi^{2} + \lambda\frac{\phi^{6}}{6!},
\ee
where $\lambda$ is a negative dimensional coupling constant.  The
free scalar field is usually approached by expanding the field as
a sum over 3-momenta of independent harmonic oscillators.  The
starting point for such a treatment is the Klein-Gordon equation
and the promotion of classical fields to operators is achieved by
the imposition of canonical commutation relations.  If we compute
the propagation amplitude from the field expansion we recover the
Green's function of the Klein-Gordon operator.  This is expected
since the canonical commutation relations are quantum mechanical
generalizations of the Poisson brackets found in classical
mechanics. Furthermore, since a point particle can be localized
with arbitrary precision, we are able to characterize such
particles as existing in well defined 3-momentum states at any
instant in time.  When we generalize the extended object
formulation of quantum mechanics we are unable to retain this
characterization because the particle is smeared out in a small
region of spacetime.  In order to incorporate the smearing in the
time direction into our infinite dimensional generalization, we
have to characterize a fuzzy particle by its 4-momentum or its
mass (when on shell).  Moreover, the commutation relations that
arise in the extended object formulation cannot be viewed as
simple quantum mechanical generalizations of the classical Poisson
brackets since the Compton wavelength of classical particles
vanishes.  It is to be noted that any attempt to impose the
noncanonical commutation relations in the usual approach will
violate Lorentz invariance since the commutator between the field
and the momentum density as a function of 3-momentum:
\be
\left[\phi({\bf p}),\pi({\bf q})\right] =
ie^{-p^{2}/m^{2}}\delta^{(3)}({\bf p} - {\bf q}),
\ee
where $p^{2} = {\bf p}^{2}$ will not be a Lorentz invariant
quantity and which in turn leads to acausal behavior in the
theory.  The reason for this behavior is because such a
generalization introduces fuzziness in space but not in the time
direction: for example,  consider the instantaneous annihilation
of a particle of finite extent.  As a consequence, we adopt a
unified, Lorentz invariant, spacetime generalization of the
extended object formulation to fields given by
\be
\left[\phi(p),\pi(q)\right] = ie^{-p^{2}/m^{2}}\delta^{(4)}(p - q),
\ee
where $p^{2} = p_{\mu}p^{\mu}$ is a Euclidean scalar.  This
generalization allows us to characterize the fuzzy particle
intermediate states as well defined 4-momentum states or as a
collection of masses (when on shell).  We are now in a position to
write down the field expansion for $\phi(x)$,
\be
\label{f-eqn} \phi(x) = \int
\frac{d^{4}p}{(2\pi)^{4}}\frac{1}{\sqrt{p^{2} +
m^{2}}}(a_{p}e^{-ipx} + a^{\dagger}_{p}e^{ipx}),
\ee
and the corresponding momentum density,
\be
\pi(x) = \int \frac{d^{4}p}{(2\pi)^{4}}(-i)\sqrt{p^{2} +
m^{2}}(a_{p}e^{-ipx} - a^{\dagger}_{p}e^{ipx}).
\ee
 The weights in the above equations have been chosen so that,
 with canonical commutations underlying,
$\left[a_{p},a^{\dagger}_{q}\right] = (2\pi)^{4}\delta^{(4)}(p - q)$,
we obtain
\be
\label{up-eqn} D_{R}(x - y) = \langle 0|\phi(x)\phi(y)|0\rangle =
\int \frac{d^{4}p}{(2\pi)^{4}}\frac{1}{p^{2} + m^{2}}e^{-ip(x -
y)},
\ee
which is the retarded scalar field propagator or the Green's
function for the Klein-Gordon operator.  We observe that
Eq.~(\ref{up-eqn}) is symmetric under $x\leftrightarrow y$ (just
change $p \rightarrow -p$) implying that
$\left[\phi(x),\phi(y)\right] = 0$ for all $x$ and $y$.The
vanishing of this commutator can be explained by noting that
$\phi(x)$ is a field which creates and destroys 4-momentum (or
mass) states and the measurement of such a field at one spacetime
point cannot affect the measurement at another spacetime point
since 4-momentum states are relativistic invariants.  However,
there is a nonzero propagation amplitude for 4-momentum states.
Let us denote by $\phi^{(3)}(x)$ the usual 3-momentum expansion of
the field obtained by quantizing the classical equation of motion,
that is,
\be
\label{3m-eqn} \phi^{(3)}(x) = \int
\frac{d^{3}p}{(2\pi)^{3}}\frac{1}{\sqrt{2E_{{\bf
p}}}}(a_{p}e^{-ipx} + a^{\dagger}_{p}e^{ipx}),
\ee
where $E_{{\bf p}} = \sqrt{{\bf p}^{2} + m^{2}}$.  Then from
Eq.~(\ref{up-eqn}) it follows that
\be
D_{R}(x - y) = \langle 0|\phi(x)\phi(y)|0\rangle = \theta(x^0 -
y^0)\langle 0|[\phi^{(3)}(x),\phi^{(3)}(y)]|0\rangle.
\ee
Thus, the 4-momentum expansion generates the correct propagation
amplitude for the scalar field which admits a representation as
the vacuum expectation value of the commutator of 3-momentum field
expansions.  Since the ordinary scalar field is quantized by
generalizing the spatial commutation relations obtained from
ordinary quantum mechanics to infinite dimensions, time remains a
parameter, and questions of causality arise.  It is well known
that this commutator vanishes over spacelike intervals and that
causality is preserved in the ordinary scalar field.  This is
expected since the propagation amplitude for the ordinary scalar
field can also be obtained from a Lorentz invariant, spacetime
generalization of a causal structure, namely, ordinary quantum
mechanics. When we impose noncanonical commutation relations
$\left[a_{p},a^{\dagger}_{q}\right] =
(2\pi)^{4}e^{-p^{2}/m^{2}}\delta^{(4)}(p - q)$ on the field and
momentum density expansions we obtain the  propagation amplitude
describing intermediate states as:
\be
\label{pamp-eqn}
D(x - y) = \langle 0|\phi(x)\phi(y)|0\rangle =
\int \frac{d^{4}p}{(2\pi)^{4}}\frac{e^{-p^{2}/m^{2}}}{p^{2} +
m^{2}}e^{-ip(x - y)}.
\ee
We once again observe the symmetry under $x \leftrightarrow y$ (by
changing $p \rightarrow -p$) which reflects the fact that
$\phi(x)$ creates and destroys relativistically invariant
4-momentum states.  The correspondence between Euclidean space and
Minkowski space propagation amplitudes is effected by means of the
variable transformation $p_0 \rightarrow ip_0$ which has been
mentioned before. This cannot be viewed as a rotation of the
contour in the sense of the Wick rotation since the contour of
integration does not close in this case.  We note that the
corresponding $dp_0$ integration limits in Minkowski space go from
$-i\infty$ to $i\infty$.  In order to explain this feature,
consider the commutation relations arising from the quantum
mechanics of extended objects
\be
\label{mi-eqn}
\left[ X_{f_i},P_j\right] =
ie^{-P^2/m^2}\delta_{ij}, \mbox{ and }\left[X_{f_0},P_0\right] =
-ie^{-P_0^2/m^2},
\ee
where $P^2 = P_iP^i$.  The spacetime generalization of these
commutation relations in Minkowski space can be achieved by
writing them as
\be
\left[X_{f_{\mu}},P_{\nu}\right] = i\eta_{\mu\nu}e^{-P^2/m^2},
\ee
where $P^2 = P_{\mu}P^{\mu}$ is a Lorentz scalar and
$\eta_{\mu\nu} = (-1,1,1,1)$.  However, such spacetime commutation
relations do not lend themselves to a Lorentz invariant
generalization to fields due to the presence of the Minkowski
metric.  The only other possibility is to re-express the fuzzy
time and energy commutation relation as
\be
\left[iX_{f_0},iP_0 \right] = ie^{-P_0^2/m^2}
\ee
in order to get rid of the relative negative sign in
Eq.~(\ref{mi-eqn}).  This would imply that the $dp_0$ integration
limits would now go from $-i\infty$ to $i\infty$ and it is for
this reason that the Minkowski space propagation amplitude remains
bounded in spite of the positive term in the exponential.  If the
contour of integration were to close this would be a Wick rotation
as is the case in the limit of vanishing Compton wavelength.
Equivalently, we can formulate the theory in Euclidean space.  The
Euclidean momentum space propagator is given by:
\be
\label{prop-eqn}
{\tilde D(p)} = \frac{e^{-p^{2}/m^{2}}}{p^{2} + m^{2}},
\ee
which is bounded from above and below since $p^{2}$ is a Euclidean scalar.  In the limit as the Compton wavelength vanishes we recover the usual scalar field propagator (Euclidean).  We note that in arriving at this propagator we have employed a different quantization prescription but we have not changed the Lagrangian, that is, the classical field theory is left undisturbed.  Since the quantization prescription arises from the quantum mechanics of extended objects which has no classical counterpart, we are justified in employing this procedure.  A crucial feature of this propagator is the Gaussian damping which renders the theory finite; for example, insertion of an arbitrary number of vertices in the N-point function of a $\phi^{6}$ theory would no longer make the graphs divergent because the Gaussian damping eliminates the high frequency modes.  Since the extended object formulation is a causal structure its infinite dimensional, Lorentz invariant, generalization will obey causality.  In order to prove that causality is not violated we need to establish that the propagation amplitude for 4-momentum states given in Eq.~(\ref{pamp-eqn}) does not admit a representation as the vacuum expectation value of the commutator of two 3-momentum field expansions.  This is crucial since the propagation amplitude does not vanish over spacelike intervals but this in itself does not violate causality.  The correct statement of causality requires us to determine whether a measurement made at one spacetime point can affect a measurement made at another spacetime point outside the light cone.  Causality is preserved if the commutator of two 3-momentum field expansions vanishes over spacelike intervals {\it provided $\phi^{(3)}(x)$ exists}.  If $\phi^{(3)}(x)$ does not exist the nonvanishing of the propagation amplitude over spacelike intervals will have no implication for causality.  In scalar field theory with canonical commutation relations underlying we are able to construct field expansions $\phi^{(3)}(x)$ for intermediate states which are sum over 3-momenta of independent harmonic oscillators obtained by simply quantizing the Klein-Gordon equation.  We observe that with noncanonical commutation relations underlying we can no longer construct $\phi^{(3)}(x)$ because well defined 3-momentum characterizations for fuzzy particle states do not exist which we expect from considerations of Lorentz invariance and causality.  In order to prove that it is impossible to construct well defined 3-momentum states for fuzzy particles, assume that there exists $\phi^{(3)}(x)$ such that
\be
\label{phi3-eqn}
\theta(x^0 - y^0)\langle
0|[\phi^{(3)}(x),\phi^{(3)}(y)]|0\rangle =
\int\frac{d^{4}p}{(2\pi)^{4}}\frac{e^{-p^{2}/m^{2}}}{p^{2} +
m^{2}}e^{-ip(x - y)}.
\ee
In particular, for $x = y$, we require that
\be
\label{int-eqn}
\int{d^{4}p}\frac{e^{-p^{2}/m^{2}}}{p^{2} + m^{2}} = 0.
\ee
The Euclidean momentum space integral can be easily evaluated by
employing the variable transformation $p^2 = u$ but in order to
demonstrate a method which will be useful in proving unitarity, we
adopt a different approach.  Due to the essential singularity at
infinity the $dp_0$ integral cannot be evaluated by contour
integration.  Switching to dimensionless variables we wish to
evaluate
\be
\int_{-\infty}^{\infty}dx\frac{e^{-x^{2}}}{x^{2} + x_{0}^{2}} =
\frac{1}{2ix_{0}}\left[\int_{-\infty}^{\infty}dx\frac{e^{-x^{2}}}{x
- ix_{0} + i\epsilon} -
\int_{-\infty}^{\infty}dx\frac{e^{-x^{2}}}{x + ix_{0} +
i\epsilon}\right].
\ee
So the problem is reduced to finding
\be
f(x_{0}) = \int_{-\infty}^{\infty}dx\frac{e^{-x^{2}}}{x - ix_{0} +
i\epsilon},
\ee
which defines the so called Hilbert transform.  Now
\bea
\frac{df(x_{0})}{dx_{0}}
&=& \int_{-\infty}^{\infty}dx
    e^{-x^{2}}\frac{d}{dx_{0}}\frac{1}{x - ix_{0} +
    i\epsilon}\nonumber \\
&=& -i\int_{-\infty}^{\infty}dx
    e^{-x^{2}}\frac{d}{dx}\frac{1}{x - ix_{0} +i\epsilon}\nonumber\\
&=& i\int_{-\infty}^{\infty}dx\frac{1}{x - ix_{0} +
    i\epsilon}\frac{d}{dx}e^{-x^{2}}\\ &=& -i\int_{-\infty}^{\infty}dx
    e^{-x^{2}}\frac{2x}{x - ix_{0} + i\epsilon}\nonumber \\
&=& -2i\int_{-\infty}^{\infty}dx e^{-x^{2}}\left[1 +
    \frac{ix_{0}}{x - ix_{0} + i\epsilon}\right]\nonumber\\
&=& -2i\sqrt{\pi} + 2x_{0}f(x_{0})\nonumber.
\eea
If we multiply this differential equation by $e^{-x_{0}^{2}}$ and
integrate from $0$ to $x_{0}$ using the initial value $f(0) =
i\pi$ obtained by simple contour integration we get
\be
e^{-x_{0}^{2}}f(x_{0}) = i\pi - 2i\sqrt{\pi}\int_{0}^{x_{0}}dx'e^{-x'^{2}}.
\ee
If we define $erf(x_{0}) = \frac{2}{\sqrt{\pi}}\int_{0}^{x_{0}}dy
e^{-y^{2}}$ we can express the solution as
\be
f(x_{0}) = i\pi e^{x_{0}^{2}}[1 - erf(x_{0})].
\ee
Using this formula we can evaluate the 4-momentum integral to obtain
\be
\label{ab-eqn}
\int d^{4}p\frac{e^{-p^{2}/m^{2}}}{p^{2} + m^{2}} =
\frac{-2\pi e}{m^2}\int\frac{d^{3}p}{2E_{{\bf p}}}erf(E_{{\bf
p}}),
\ee
where $E_{\bf p} = \sqrt{{\bf p}^{2} + m^{2}}$.  The 3-momentum
integral over the error function is a nonzero quadrature and we
arrive at a contradiction to our original requirement that the
integral in Eq.~(\ref{int-eqn}) vanish.  As the Compton wavelength
vanishes ($\frac{1}{m} \rightarrow 0$) the left hand side of
Eq.~(\ref{ab-eqn}) reverts to the ordinary scalar field propagator
(for $x = y$) and the right hand side vanishes, a result we
expect. Hence, the existence of 3-momentum characterizations for
fuzzy particle states would have been an indication of acausal
behavior.  Similarly, we can demonstrate that there exists no
$\phi^{(3)}(x)$ which can satisfy Eq.~(\ref{phi3-eqn}) for $x \neq
y$.  Evaluation of the propagation amplitude for $x \neq y$ leads
to a sum of error functions of arguments involving $x - y$ and a
functional form for $\phi^{(3)}(x)$ as a sum of creation and
annihilation operators becomes impossible to obtain.  Thus, the
fact that we have constructed the field theory from a Lorentz
invariant, spacetime generalization of a causal formulation,
namely, the quantum mechanics of extended objects, combined with
the fact that 3-momentum characterizations do not exist for fuzzy
particle intermediate states allows us to conclude that the
statement of causality is not violated in the theory.
\section{unitarity in the theory}
The optical theorem relates the forward scattering amplitude to
the total cross section for production of all final states.  Since
the imaginary part of the forward scattering amplitude gives the
attenuation of the forward going wave as the beam passes through
the target, it is natural that this quantity should be
proportional to the probability of scattering\cite{peskin}.  In
the quantum field theory of point particles each diagram
contributing to an S-matrix element is purely real unless some
denominators vanish so that the $i\epsilon$ prescription for
treating the poles becomes relevant, that is, a Feynman diagram
yields an imaginary part for the scattering amplitude ${\cal M}$
only when the virtual particles in the diagram go on mass shell.
In the quantum field theory of extended objects because of the
Gaussian damping term in the propagator the scattering amplitude
exhibits an imaginary part even below the threshold for production
of multiparticle states.  However, this is the unphysical part of
the scattering amplitude and does not contribute to the optical
theorem since we only need to verify whether the physical
scattering amplitude which lies above the threshold for the
production of multiparticle states is equal to the total cross
section for the production of all such states.  Let us first
calculate the scattering amplitude at order-$\lambda^{2}$ with a
$\phi^{6}$ interaction.  The order-$\lambda^{2}$ diagram is shown
in figure 1 and the  contribution of this diagram is given by the
Lorentz invariant integral
\be
i\delta{\cal M} =
\frac{\lambda^{2}}{2}\int\frac{d^{4}q}{(2\pi)^{4}}\frac{
e^{-(\frac{k}{2} + q)^{2}/m^{2}}} {(\frac{k}{2} + q)^{2} +
m^{2}}\frac{e^{-(\frac{k}{2} - q)^{2}/m^{2}}}{(\frac{k}{2} -
q)^{2} + m^{2}},
\ee
where $k = \sum_{i=1}^{4}k_{i}$ and we have chosen a symmetric
routing of momenta.  We observe that the because of the
dimensional coupling (mass dimension = -2) we would need to
increase the powers of momentum in the integrand in order to
render the graph dimensionless, but because of the Gaussian
damping factors the graphs will not diverge at any order.  Due to
the essential singularities in the integrand arising due to the
Gaussian damping terms in the propagators, this integral cannot be
evaluated by contour integration and for the same reason the
Cutkosky cutting rules do not apply.  Consequently, we employ the
standard method of Feynman parameters to obtain
\be
i\delta{\cal M} =
\frac{\lambda^{2}}{16\pi^2}\int_{0}^{1}dxe^{-k^{2}/2m^{2}}\int_{0}^{\infty}dl\,
l^{3}\frac{e^{-2l^{2}/m^{2} - 2lk/m^{2}}}{(l^{2} + \Delta)^{2}},
\ee
where $l = q + \frac{kx}{2} - \frac{k(1 - x)}{2}$ and $\Delta =
k^{2}x(1 - x) + m^{2}$.  We note that since we are dealing with
Euclidean momenta, a Wick rotation would be superfluous.  The
imaginary part shows up in the $dl$ integration which can be
recast as
\be
\label{l'-eqn}
\left. \frac{d}{d\Delta}
\frac{d^{3}}{d\alpha^{3}}
e^{-\frac{k^{2}}{2m^{2}}(1 - \alpha^{2})}\int_{0}^{\infty}dl
\frac{e^{-\frac{2}{m^{2}}
(l + \frac{\alpha k}{2})^{2}}}{(l^{2} + \Delta)}
\right|_{\alpha = 1}.
\ee
The integral can be evaluated by the formula developed in the
previous section and noting that Eq.~(\ref{l'-eqn}) contains a
shifted Gaussian and that  the integration limits are from $0$ to
$\infty$ we obtain
\bea
\label{erf-eqn}
\lefteqn{ i\delta{\cal M} =
\frac{\lambda^2}{16\pi^{2}}(\frac{m^{2}}{2k})^{3} \int_{0}^{1} dx
\frac{d}{d\Delta}\frac{d^{3}}{d\alpha^{3}}
e^{-\frac{k^{2}}{2m^{2}}(1 - \alpha^{2})} \times } \\ & & \left.
\left[\frac{1}{2i\sqrt{\Delta}} \left( h(k) -
i\frac{\pi}{2}erfc(\frac{\alpha k}{m\sqrt{2}}) erf(\frac{i\alpha
k}{m\sqrt{2}} - \sqrt{\Delta})\right) e^{({i\alpha k}/{m\sqrt{2}}
- \sqrt{\Delta})^2}
 + (\sqrt{\Delta} \,\leftrightarrow \, -\sqrt{\Delta})\right]
\right|_{\alpha = 1}, \nonumber
\eea
where
\be
h(k) = \left(\int_{\frac{\alpha k}{m\sqrt{2}}}^{\infty}\!\! + \:
e^{-\alpha^2k^2/2m^2}\! \int_{\frac{i\alpha
k}{m\sqrt{2}}}^{\sqrt{\Delta}}\right)dx \frac{e^{-x^{2}}}{x},
\ee
and $erfc$ is the complementary error function.  Consider the
analytic structure of the scattering amplitude.  The square root
of $\Delta$ has a branch cut when its argument becomes negative,
that is, when
\be
k^{2}x(1 - x) + m^{2} < 0.
\ee
The product $x(1 - x)$ is at most $\frac{1}{4}$ so ${\cal M}(k)$
has a branch cut beginning at
\be
k^{2} = -4m^{2}
\ee
at the threshold for the creation of a multiparticle state. When
$\Delta < 0$ (that is, $-k^{2} > 4m^{2}$) $h(k)$, and the
exponential factors in Eq.~(\ref{erf-eqn}) become purely real, and
since
\be
erf(ix) = \frac{i}{\sqrt{2\pi}}\int_{0}^{x}e^{x'^{2}}dx',
\ee
the error functions with complex arguments in Eq.~(\ref{erf-eqn})
become purely imaginary functions.  Consequently, the right hand
side of Eq.~(\ref{erf-eqn}) becomes purely real.  Due to the extra
factor of $i$ in the scattering amplitude we find that ${\cal
M}(k)$ is purely imaginary above threshold, that is,
\be
Im {\cal M}(k) = {\cal M}(k) \;\mbox{when}\; \Delta < 0,
\ee
a fact which will be important in proving unitarity.  This fact
implies that the forward going wave is purely attenuating as the
beam passes through the target and the cross section which is
proportional to the probability of scattering is at its maximum.
Thus, the hitherto nonrenormalizable interaction saturates the
{\it unitarity bound}.  This bound is saturated whenever the
scattering phase shift is an odd multiple of $\frac{\pi}{2}$.
This condition implies the formation of resonances or metastable
bound states.  In a scattering process at a resonant energy, the
incident particle has a larger probability of becoming temporarily
trapped in such a metastable bound state and this possibility
increases the scattering cross section.  Resonances are not
observed in $\phi^{4}$ theory scattering processes, at least at
this order, and their appearance in $\phi^{6}$ theory may reflect
the increased interaction field strength.

Below the threshold for production of multiparticle states, that
is, when $\Delta > 0$ we observe that the scattering amplitude has
an imaginary part as we expect.  This is the unphysical part of
the scattering amplitude since the virtual particles cannot go on
shell and it does not contribute to the scattering process.  By
cutting through the diagram as shown in figure 1 we can evaluate
the cross section which has the value
\bea
\lefteqn{\int e^{q_{1}^{2}/m^{2}}\frac{d^{4}q_{1}}
{(2\pi)^{4}}e^{q_{2}^{2}/m^{2}}
\frac{d^{4}q_{2}}{(2\pi)^{4}}
{\cal M}^{*}(p_{1}p_{2}p_{3}p_{4}  \rightarrow q_{1}q_{2})
\times } \nonumber \\
& & \hspace{2in}
{\cal M}(k_{1}k_{2}k_{3}k_{4} \rightarrow q_{1}q_{2})
\delta^{(4)}(k_{1} + k_{2} + k_{3} + k_{4} - q_{1} - q_{2})
\eea
times an overall delta function $(2\pi)^{4}\delta^{(4)}(\sum_{i =
1}^{4}k_{i} - \sum_{i = 1}^{4}p_{i})$.  The commutation relations
$\left[a_{p},a^{\dagger}_{q}\right] =
(2\pi)^{4}e^{-p^{2}/m^{2}}\delta^{(4)}(p - q)$ modify the
completeness relation as shown.  To see this, we observe that
\be
\int e^{p^{2}/m^{2}}\frac{d^{4}p}{(2\pi)^{4}}\langle q|p\rangle\langle p|
= \langle q|
\ee
since
\be
\langle q|p\rangle = \langle 0|a_{q}a^{\dagger}_{p}|0\rangle =
(2\pi)^{4}e^{-p^{2}/m^{2}}\delta^{(4)}(p - q)
\ee
implying that
\be
\int e^{p^{2}/m^{2}}\frac{d^{4}p}{(2\pi)^{4}}|p\rangle\langle p| = 1.
\ee
We note that the normalization we have chosen for fuzzy particle
intermediate states is Lorentz invariant.  The nontrivial
contribution to the S-matrix comes from the term:
\be
\label{ma-eqn}
\langle{\bf k_{1}k_{2}k_{3}k_{4}}|\phi^{6}|q_{1}q_{2}\rangle.
\ee
In contracting the $\phi$ operators with the external 3-momentum
states we make use of the usual expansion $\phi^{(3)}(x)$ given in
Eq.~(\ref{3m-eqn}) to obtain
\be
\phi^{(3)}(x)|{\bf p}\rangle = e^{-ipx}|0\rangle.
\ee
In contracting the $\phi$ operators with fuzzy particle
intermediate states we avail the 4-momentum expansions to obtain
\bea
\phi(x)|p\rangle &=& \int \frac{d^{4}k}{(2\pi)^{4}}\frac{1}{\sqrt{k^{2} + m^{2}}}(a_{k}e^{-ikx}a^{\dagger}_{p}|0\rangle) \nonumber \\
&=& \frac{e^{p^{2}/m^{2} - ipx}}{\sqrt{p^{2} + m^{2}}}.
\eea
By making use of these contractions we can evaluate the matrix
element given in Eq.~(\ref{ma-eqn}).  The value of the cut diagram
becomes
\be
6!(\frac{-i\lambda}{6!})\frac{e^{-q_{1}^{2}/m^{2}}}{\sqrt{q_{1}^{2}
+ m^{2}}}\frac{e^{-q_{2}^{2}/m^{2}}}{\sqrt{q_{2}^{2} +
m^{2}}}(2\pi)^{4}\delta^{(4)}(k_{1} + k_{2} + k_{3} + k_{4} -
q_{1} - q_{2}).
\ee
This is exactly of the form $i{\cal M}(2\pi)^{4}\delta^{(4)}(k_{1}
+ k_{2} + k_{3} + k_{4} - q_{1} - q_{2})$ with
\be
{\cal M} = -\lambda\frac{e^{-q_{1}^{2}/m^{2}}}{\sqrt{q_{1}^{2} + m^{2}}}\frac{e^{-q_{2}^{2}/m^{2}}}{\sqrt{q_{2}^{2} + m^{2}}}.
\ee
An identical computation leads to the same value for the complex
conjugate and hence we obtain the cross section(without the
kinematical factors) as
\be
\lambda^{2}\int d^{4}q_{1}d^{4}q_{2}\,\frac{e^{-q_{1}^{2}/m^{2}}}{q_{1}^{2} + m^{2}}\frac{e^{-q_{2}^{2}/m^{2}}}{q_{2}^{2} + m^{2}}(2\pi)^{4}\delta^{(4)}(k_{1} + k_{2} + k_{3} + k_{4} - q_{1} - q_{2}).
\ee
This is exactly twice the scattering amplitude $2{\cal M}(k)$ for
the whole process.  We have already proved that above the
threshold for production of multiparticle states
\be
Im {\cal M}(k) = {\cal M}(k), \;\mbox{when}\; \Delta < 0.
\ee
Thus, it follows that the imaginary part of the physical scattering amplitude is equal to the total cross section for production of all final states after the requisite kinematical factors needed to build a cross section have been supplied.  Therefore, the optical theorem is obeyed at order-$\lambda^{2}$ in perturbation theory and unitarity is preserved at this order.

We now proceed to evaluate the optical theorem at order-$\lambda^{4}$ in perturbation theory.  Consider a typical order-$\lambda^{4}$ diagram as shown in figure 2.  The value of this Feynman diagram is
\bea
\lefteqn{
i\delta{\cal M} = \frac{\lambda^{4}}{2(4!)^{2}}
\int\,\prod_{1 = 1}^{4}\frac{d^{4}q_{i}}{(2\pi)^{4}} \frac{d^4
r_i}{(2\pi)^4} \delta^{(4)}(k - \sum_{i = 1}^{4}q_{i} - \sum_{i =
1}^{4}r_{i}) \times } \\ & & \hspace{1in} \left[\frac{e^{-\sum_{1
= 1}^4 q_i^2/m^2}} {\prod_{i = 1}^{4}(q_{i}^{2} + m^{2})}
\frac{e^{-(k_{1} - \sum_{i = 1}^{4}q_{i})^{2}/m^{2}}} {(k_{1} -
\sum_{i = 1}^{4}q_{i})^{2} + m^{2}} \frac{e^{-\sum_{1 =
1}^{4}r_{i}^{2}/m^{2}}} {\prod_{i = 1}^{4}(r_{i}^{2} + m^{2})}
\frac{e^{-(p_{2} - \sum_{i = 1}^{4}r_{i})^{2}/m^{2}}} {(p_{2} -
\sum_{i = 1}^{4}r_{i})^{2} + m^{2}}\right], \nonumber
\eea
where $k = k_{1} + k_{2}$.  Suppose we express
\bea
q_{1} &=& q\sin\omega\sin\theta\cos\phi =
qf_{1}(\omega,\theta,\phi)\nonumber\\ q_{2}
&=&
q\sin\omega\sin\theta\sin\phi = qf_{2}(\omega,\theta,\phi)\\ q_{3}
&=& q\sin\omega\cos\theta = qf_{3}(\omega,\theta)\nonumber\\ q_{4}
&=& q\cos\omega = qf_{4}(\omega)\nonumber
\eea
and similarly for the $r_{i}$.  Introducing these four dimensional
spherical coordinates and evaluating the delta function we obtain
\bea
\label{m-eqn}
\lefteqn{i\delta{\cal M} =
\frac{\lambda^{4}(2\pi^{2})^{8}(m^2)^{12}}{2(4!)^{2}(2\pi)^{32}}
\frac{d^{6}}{d\alpha^{6}}\frac{d^{6}}{d\beta^{6}}
 \left\{ \int\,d^{4}q \left(\prod_{i = 1}^{4} f_i\right)^6\,
\frac{e^{-\alpha q^{2}/m^{2}}}
{\prod_{i = 1}^{4}(q^{2}f_{i}^{2} + m^{2})}
\right. \times} \\
& & \hspace{1in}
\left. \frac{e^{-(k_{1} - qf)^{2}/m^{2}}}{(k_{1} - qf)^{2} + m^{2}}
\frac{e^{-\beta(\frac{k}{f} - q)^{2}/m^{2}}}
{\prod_{i = 1}^{4}(\frac{k}{f} - q)^{2}f_{i}^{2} + m^{2})}
\left. \frac{e^{(p_{2} - k + qf)^{2}/m^{2}}}
{(p_{2} - k + qf)^{2} + m^{2}}\right|_{\alpha,\beta = 1} \right\}, \nonumber
\eea
where $f = \sum_{i = 1}^{4}f_{i}$.  Employing the method of
Feynman parameters we can re-express the denominators of
Eq.~(\ref{m-eqn}) in terms of
\be
D = q^{2}\sum_{i = 1}^{4}x_{i}f_{i}^{2} + x_{5}(k - qf)^{2} +
\sum_{j = 6}^{9}x_{j}f_{j}^{2}(k/f - q)^{2} + x_{10}(p_{2} - k + qf)^{2}
+ m^{2} + i\epsilon,
\ee
where $f_{j} = f_{i}$.  We observe that $D$ can be written as
\be
\label{q-eqn}
D = C_{1}q^{2} + C_{2}q + C_{3} + m^{2} ,
\ee
where the $C_{i}, i = 1,2,3$ are angle dependent coefficients given by:
\bea
C_{1} &=& \sum_{1 = 1}^{4}x_{i}f_{i}^{2} + x_{5}f^{2} + \sum_{j =
6}^{9}x_{j} f_{j}^{2} + x_{10}f^{2} \nonumber\\ C_{2} &=&
-2x_{5}kf - \frac{2k}{f}\sum_{j = 6}^{9}x_{j}f_{j}^{2} +
2x_{10}p_{2}f - 2x_{10}kf\\ C_{3} &=& k^{2}\left(x_{5} +
\frac{1}{f^{2}}\sum_{j = 6}^{9}x_{j} f_{j}^{2} + x_{10}\right) +
x_{10}\left(p_{2}^{2} - 2p_{2}k\right). \nonumber
\eea
For a scattering process $q^{2} > 0$ implying that $m^{2} + C_{3}
- \frac{C_{2}^{2}}{4C_{1}} < 0$.  Eq.~(\ref{q-eqn}) motivates us
to express the denominator as
\be
D = l^{2} + \Delta ,
\ee
where $l = \sqrt{C_{1}}q + \frac{C_{2}}{2\sqrt{C_{1}}}$ and $\Delta = m^{2}
 + C_{3} - \frac{C_{2}^{2}}{4C_{1}} < 0$ (for scattering).
Therefore, the fourth order scattering amplitude can be recast as
\bea
\lefteqn{ i\delta{\cal M} =
\frac{\lambda^{4}(2\pi^{2})^{8}(m^2)^{12}}{2(4!)^{2}9!(2\pi)^{32}}
e^{-(k_{1}^{2} + p_{2}^{2} + 2p_{2}k)/m^{2}}\times } \nonumber \\
& & \hspace{0.5in} \int_{0}^{1}\,dx_{1}\cdots dx_{10}
\delta(\sum_{i = 1}^{10}x_{i} - 1)
\frac{d^{9}}{d\Delta^{9}}\frac{d^{6}}{d\alpha^{6}}
\frac{d^{6}}{d\beta^{6}}\int d^{4}l\left(\prod_{i = 1}^{4}
f_i\right)^6 C_1^2 \left. \frac{e^{-g(l)/m^2}} {l^2 +
\Delta}\right|_{\alpha, \beta = 1},
\eea
where
\be
g(l) = (l/\sqrt{C_{1}} - C_{2}/2C_{1})^{2}
(\alpha + 2f^{2} + \beta) - (l/\sqrt{C_{1}} -
C_{2}/2C_{1})(2k_{1}f + \frac{2\beta k}{f} + 2kf).
\ee
Apart from angular coefficients $g(l)$ is a function only of
$l^{2}$ and $l$.  Therefore, the $dl$ integration can be brought
into the same form as in Eq.~(\ref{l'-eqn}) by collecting the
angular coefficients.  The results of such an integration leads to
a sum of error functions of complex arguments each of which is
multiplied by an exponential factor also of complex argument as in
Eq.~(\ref{erf-eqn}).  Since the subsequent angular integration
does not change the analytic structure of the scattering amplitude
we can conclude (as before) that
\be
\label{im-eqn} {\rm Im} {\cal M}(k) = {\cal M}(k), \;\mbox{when}\;
\Delta < 0.
\ee
Thus, the physical scattering amplitude saturates the unitarity
bound and resonances are found to occur at order-$\lambda^{4}$ in
perturbation theory. By cutting through the diagram as shown in
figure 2 and evaluating the cut diagrams we can compute the cross
section.  A simple calculation shows that the cross section is
proportional to
\bea
\lefteqn{ \frac{\lambda^{4}}{(4!)^{2}} \int\,\prod_{1 =
1}^{4}\frac{d^{4}q_{i}}{(2\pi)^{4}} \frac{d^4 r_i}{(2\pi)^4}
\delta^{(4)}(k - \sum_{i = 1}^{4}q_{i} - \sum_{i = 1}^{4}r_{i})
\times \frac{e^{-\sum_{1 = 1}^4 q_i^2/m^2}} {\prod_{i =
1}^{4}(q_{i}^{2} + m^{2})} \times } \nonumber \\ & & \hspace{1cm}
\frac{e^{-(k_{1} - \sum_{i = 1}^{4}q_{i})^{2}/m^{2}}} {(k_{1} -
\sum_{i = 1}^{4}q_{i})^{2} + m^{2}} \frac{e^{-\sum_{1 =
1}^{4}r_{i}^{2}/m^{2}}} {\prod_{i = 1}^{4}(r_{i}^{2} + m^{2})}
\frac{e^{-(p_{2} - \sum_{i = 1}^{4}r_{i})^{2}/m^{2}}} {(p_{2} -
\sum_{i = 1}^{4}r_{i})^{2} + m^{2}} ,
\eea
which is exactly twice the scattering amplitude for the whole
process.  By making use of Eq.~(\ref{im-eqn}) we find that the
imaginary part of the scattering amplitude is equal to the total
cross section for production of all final states after the
requisite kinematical factors needed to build a cross section have
been supplied.  Thus, the optical theorem holds at
order-$\lambda^{4}$ in perturbation theory and unitarity is
preserved at this order.  Since the dynamical contributions to the
scattering process begin at order-$\lambda^{4}$ we can expect the
optical theorem to hold at every order in perturbation theory and
by introducing $d$ dimensional coordinates ($d > 4$) we can
similarly verify the optical theorem at higher orders in
perturbation theory.

\section{The Kallen-Lehmann representation}
The Kallen-Lehmann representation gives us a spectral
representation of the propagator in the interacting picture.  We
wish to determine whether the Euclidean momentum space propagator
given in Eq.~(\ref{prop-eqn}) admits such a spectral
representation in the interacting picture.  This is important
because it allows us to physically interpret the interacting
propagator as a weighted sum of free propagation amplitudes.  To
analyze the two point function
$\langle\Omega|\phi(x)\phi(y)|\Omega\rangle$ we will insert the
identity operator as a sum over a complete set of states.  We
choose these states to be eigenstates of the 4-momentum operator
P.  The two point function becomes
\be
\langle\Omega|\phi(x)\phi(y)|\Omega\rangle =
\sum_{\lambda}\int\frac{d^{4}p}{(2\pi)^{4}}e^{p^{2}/m_{\lambda}^{2}}
\langle\Omega|\phi(x)|\lambda_{p}\rangle
\langle\lambda_{p}|\phi(y)|\Omega\rangle,
\ee
where $|\Omega\rangle$ is the interacting vacuum state,
$|\lambda_{p}\rangle$ is the eigenstate of $P$, and the sum runs
over all 4-momentum states.  Using translational invariance we can
write
\be
\label{1-eqn}
\langle\Omega|\phi(x)|\lambda_{p}\rangle =
\langle\Omega|\phi(0)|\lambda_{p}\rangle e^{-ipx},
\ee
and
\be
\label{2-eqn} |\langle\Omega|\phi(0)|\lambda_{p}\rangle|^{2} =
\frac{e^{-2p^{2}/m_{\lambda}^{2}}}{p^{2} +
m_{\lambda}^{2}}Z_{\lambda}.
\ee
The probability density to create a single particle 4-momentum
state from the free vacuum can be calculated from the field
expansions given in Eq.~(\ref{f-eqn}) and is given by
\be
|\langle p|\phi(0)|0\rangle|^{2} = \frac{e^{-2p^{2}/m^{2}}}{p^{2} + m^{2}}.
\ee
Therefore, the probability density to create a given 4-momentum
state from the interacting vacuum is given by the product of the
probability density to create a one particle state from the free
vacuum times a factor $Z_{\lambda}$ which represents the effect of
the interaction on the vacuum.  We note that in ordinary scalar
field theory which is described by point interactions the
probability of creating a 4-momentum state from the free vacuum is
zero because
\be
\int d^{4}p|\langle p|\phi(0)|0\rangle|^{2} = \int
d^{4}p\frac{1}{p^{2} + m^{2}} = 0.
\ee
This quantity is necessarily zero because a nonzero probability
for creation of a 4-momentum point particle  state from the vacuum
would violate causality.  Using the 3-momentum field expansions we
observe that
\be
\int d^{4}p|\langle p|\phi(0)|0\rangle|^{2} = \langle
0|\phi^{2}(0)|0\rangle = \langle
0|[\phi^{(3)}(0),\phi^{(3)}(0)]|0\rangle
\ee
and hence the probability must vanish.  This is due to the fact
that point particle mass states simply cannot exist, and for this
reason 4-momentum expansions are not employed in ordinary scalar
field theory even though they lead to the correct propagator.  By
inserting Eq.~(\ref{1-eqn}) and Eq.~(\ref{2-eqn}) in our
expression for the two point function we obtain
\be
\langle\Omega|\phi(x)\phi(y)|\Omega\rangle =
\int_{0}^{\infty}d\mu^{2}\frac{e^{-p^{2}/\mu^{2}}}{p^{2} + \mu^{2}
+ i\epsilon}\rho(\mu^{2})e^{-ip(x - y)},
\ee
where $\rho(\mu^{2}) = \sum_{\lambda}\delta(\mu^{2} -
m_{\lambda}^{2})Z_{\lambda}$ is a positive definite spectral
density function.  Thus, our propagator admits a Kallen-Lehmann
representation which satisfies the positivity postulates of
quantum mechanics.

\section{conclusion}
In this paper we have demonstrated that the hitherto nonrenormalizable
 scalar field with a  $\lambda \phi^6/6! $ interaction can be
 rendered finite provided we characterize the intermediate states
 as fuzzy particle states.  Such a characterization is motivated by
 generalizing the quantum mechanics of extended objects to infinite
 dimensions.  We have also demonstrated that this generalization does
 not violate causality, Lorentz invariance and unitarity (verified up
 to fourth order in the coupling constant).  Furthermore, we have
 nowhere exploited the scalar character of the field in this approach.
  This suggests that other nonrenormalizable quantum field theories
 such as quantum electrodynamics with the Pauli term or the quantum
 theory of gravity can be rendered finite by implementing this procedure.

\acknowledgements
I would like to thank E.C.G~Sudarshan and
J.R.~Klauder for valuable comments.

\newpage
\begin{figure}
\centerline{
\epsfxsize = 3.5in
\epsfbox{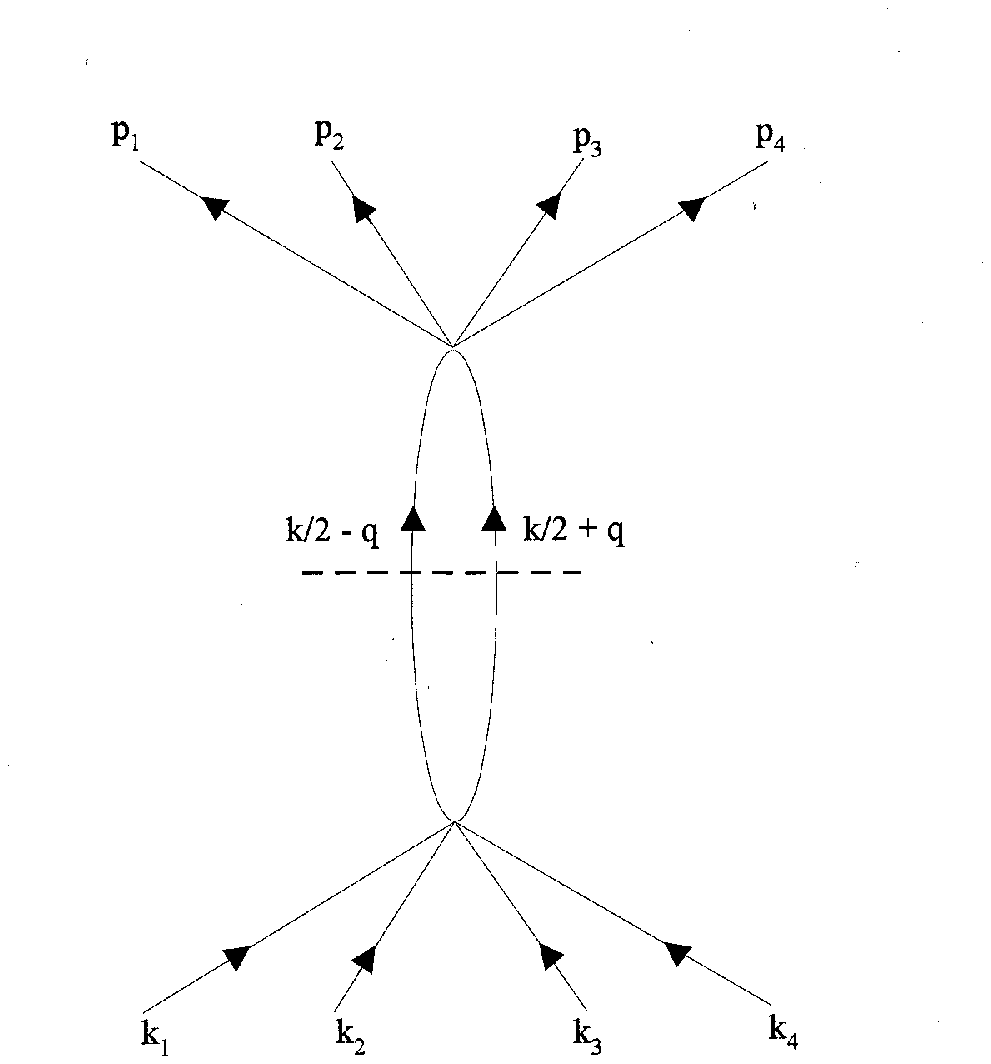}}\vskip 0.4in \hskip -1.0in
\caption{The order-$\lambda^2$ diagram in $\phi^6$ theory.}
\end{figure}

\begin{figure}
\centerline{
\epsfxsize = 3.8in
\epsfbox {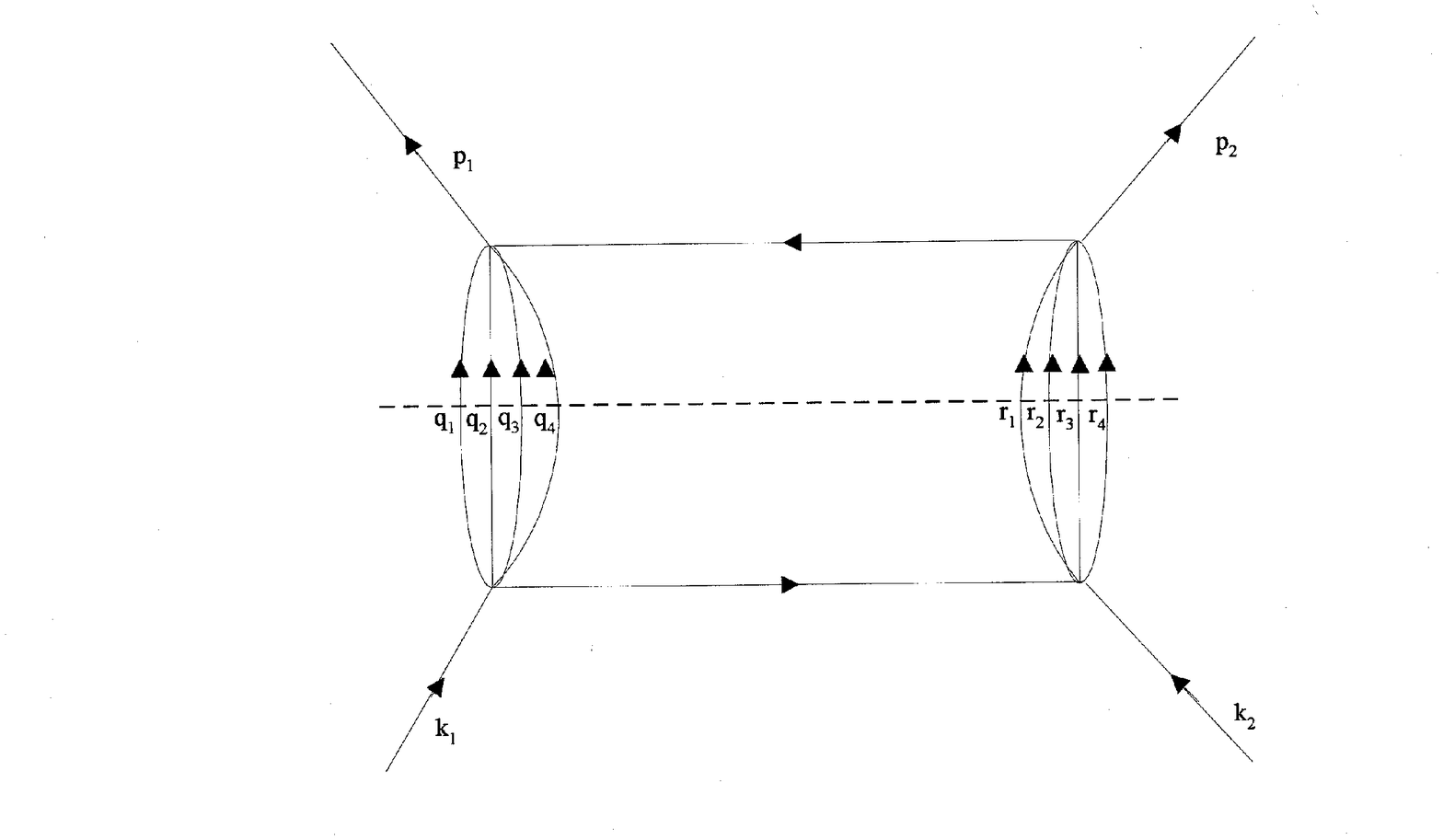}
}\vskip 0.4in
\caption{The order-$\lambda^4$ diagram in $\phi^6$ theory (the box diagram).}
\end{figure}

\end{document}